\documentclass[aps,prd,preprintnumbers,groupedaddress]{revtex4}
\usepackage{amsfonts}
\usepackage{amsmath}

\begin{document}
\preprint{hep-th/0303157, UPR-1033-T}

\title{A comment on multiple vacua, particle production and the time-dependent AdS/CFT correspondence}

\author{Vijay Balasubramanian}

\email{vijay@endive.hep.upenn.edu}

\author{Thomas S.\ Levi}
\email{tslevi@student.physics.upenn.edu}

\author{Asad Naqvi}
\email{naqvi@rutabaga.hep.upenn.edu}
\affiliation{Department of Physics and Astronomy, University of Pennsylvania, Philadelphia, PA\quad
19104-6396}

\begin{abstract}
We give an explicit formulation of the time dependent AdS/CFT correspondence when there are multiple vacua present in Lorentzian signature.
By computing sample two point functions we show how different amplitudes are related by cosmological particle production.  We illustrate our methods in two example spacetimes: (a) a ``bubble of nothing''  in AdS space, and (b) an asymptotically locally AdS spacetime with a bubble of nothing on the boundary.  In both cases the $\alpha$-vacua of de Sitter space make an interesting appearance.
\end{abstract}


\maketitle

\section{Introduction}
Recently there has been renewed interest in using the AdS/CFT correspondence to understand physics in time dependent backgrounds (e.g., \cite{birm,vijayross,mann,adstime,son02,herzog02,danielsson,esko,troost02,kraus02,nojiri}
In  Lorentzian AdS spaces the presence of both normalizable and non-normalizable fluctuations and the possibility of interesting causal structure  lead to many new phenomena (e.g., \cite{vijay981,vijay982,banks98,vijay99,giddings99,danielsson,esko,AdSbh}).  It has been shown how to compute the different types of propagators  (e.g. advanced and retarded) and correlation functions that arise in Lorentzian signature in the unique AdS vacuum \cite{danielsson,esko}.  (Also see   \cite{troost02}.)   Recent work has discussed how thermal field theory correlators can be computed using the AdS/CFT correspondence by a proper identification of boundary conditions in the corresponding AdS backgrounds which have horizons\cite{son02,herzog02}.    Finally, it has been shown that analytic continuations from the Euclidean section in different coordinate systems that cover a whole or only part of a spacetime
manifold lead to the same observable physics, even when horizons are present \cite{kraus02}.

In general, time-dependent backgrounds have multiple natural vacua.  Correlation functions in most of these vacua cannot be
obtained by analytic continuation of a Euclidean correlator.
Also, we expect non-trivial effects such as cosmological particle creation
to be apparent in the AdS/CFT correspondence. 
In interesting dynamical situations such as black hole formation from collapse, the in and out vacua that are natural before and after the black hole forms have very different characters and while the AdS/CFT correspondence naturally computes the $\langle \rm{out} | \cdots | \rm{in} \rangle$ transition amplitude,  we must compute $\langle \rm{in}| \cdots | \rm{in} \rangle$ correlators to study how the equal time correlation functions can measure horizon formation.  In this paper we show how to formulate the correspondence when a non-trivial time dependence
leads to multiple vacua. We find that properly identifying the various vacua allows us to relate correlators computed in different vacuum
states by expressions involving cosmological particle production. We also find a simple relationship between particle production in the bulk
supergravity/string theory and in the boundary gauge theory.

The paper is organized as follows. In section \ref{ads-cft} we formulate the AdS/CFT correspondence when multiple vacua are present and show how various two-point functions are related by particle production effects.   We then give two examples of situations with multiple vacua -- Sec.~\ref{rossbubble} examines the ``bubble of nothing'' in  $AdS_4$ \cite{birm,vijayross}, and Sec.~\ref{Schwars} examines the spacetime formed by placing a bubble of nothing on the boundary of $AdS_5$.
Since the surface of a bubble of nothing traces out a de Sitter space  the
 $\alpha$-vacua of de Sitter space make an interesting appearance.
In section \ref{conclusions} we present our conclusions. An appendix gives more detail for
various computations found in the main text.

\section{Multiple vacua and the AdS/CFT correspondence} \label{ads-cft}

In this section, we show how to formulate the time dependent version of the AdS/CFT correspondence  when multiple vacua are present, and compute some sample two
point functions to show how these quantities are altered by cosmological particle production.  In the original formulation of the
AdS/CFT correspondence, correlation functions in the ${\cal N}=4$ supersymmetric $SU(N)$ Yang Mills in 4d are related, in the large $N$ limit, to the bulk classical action in Euclidean $AdS_5$ \cite{maldacena97,witten98,gubser98} by
\begin{equation}
\label{euclidean ads/cft} \langle \exp \int _{\partial} \phi_0 {\cal O} \rangle_{CFT} = Z_{SUGRA} (\phi_0) \, .
\end{equation}
Here $\phi$ approaches $\phi_0$ on the boundary, $Z_{SUGRA} (\phi_0)= \exp(-I(\phi_0)) $ is the classical IIB supergravity action with the
boundary value $\phi_0$, and ${\cal O}$ is the operator dual to the field $\phi$.  Technically, correlation functions
are computed perturbatively by extending boundary values for fields (or sources in the field theory) into the AdS bulk by using a bulk-boundary propagator and then sewing these together using the usual bulk-bulk propagators to compute the AdS/CFT Feynman diagrams.  Numerous authors have explored the definition of the correspondence in Lorentzian signature (e.g., \cite{vijay981,vijay982,banks98,vijay99,giddings99,danielsson,esko} and the recent works
 \cite{son02,herzog02,troost02,kraus02}.)  We are interested in situations in which there are several natural choices of vacua.   Amplitudes and transition functions will depend on  which vacua we sandwich our operators between.   While it is clear how the choice of vacua is implemented on the CFT side of the duality, it is initially confusing how the SUGRA side, being semi-classical, contains the same data.  We will clarify this point.

In Euclidean AdS (EAdS) it is well known that specifying a field's behavior on the boundary uniquely specifies it throughout the bulk
\cite{witten98}. In contrast, due to the presence of normalizable mode solutions which can be arbitrarily added on without affecting the
boundary behavior, the solution is no longer unique (see \cite{vijay981,vijay982} for discussion).     At the level of Green
functions, in EAdS, there exists a unique bulk-boundary two point function, whereas in Lorentzian signature we have a wealth of Green
functions to choose from, all dependent on the boundary conditions (or pole prescriptions) chosen.
We expect these different Green functions to arise from grafting normalizable mode solutions onto the bulk-boundary propagator obtained by continuation from the Euclidean section.  We will see how doing this correctly allows one to select between CFT correlation functions evaluated between different vacuum states.

While we will give a prescription for treating multiple vacua in any coordinate system, it is easiest to explain the procedure by choosing a particular behaviour for the metric.   The Poincar\'{e} patch of AdS$_{d+1}$ is  given by
\begin{equation}
\label{pads metric} ds^2 = l^2/z^2 (-dt^2 +d {\bf x} ^2 + dz^2).
\end{equation}
(We will usually set $l=1$.)   In (\ref{pads metric}) the AdS boundary occurs at $z=0$ while $z=\infty$ is a coordinate horizon.   Now consider a general time-dependent solution to Einstein's equations with a negative cosmological constant which is asymptotically Poincar\'{e},
\begin{equation}
\label{modified pads metric} ds^2 \stackrel{z \rightarrow 0}{\longrightarrow}1/z^2 (dz^2 +d\tilde{s} ^2),
\end{equation}
where tilde denotes the boundary metric.   As $z \rightarrow 0$ the boundary and radial parts of the wave equation separate and the solution to the scalar wave equation will be given by $\phi(z,{\bf b})=f(z)\phi_0({\bf b})$, where ${\bf b}$ are the boundary coordinates.   As is standard, $f(z)$ has
two possible scalings (normalizable  and non-normalizable) at the boundary, namely $\lim_{z \rightarrow 0} f(z) = z^{2 h_{\pm}}$ where $2h_{\pm} = d/2 \pm \nu$ and $\nu = 1/2
\sqrt{d^2 +4m^2}$ \cite{vijay981}.

If there is a time dependence in the metric we can have multiple choices for vacuum states.  We can display this by considering different  expansions for the field $\phi$ over the normalizable modes
\begin{equation} \label{phi expansion}
\phi(z,{\bf b})= \sum_k a_k u_k (z,{\bf b}) + a_k ^\dagger u_k ^* (z,{\bf b}) = \sum_k \bar{a_k} \bar{u_k} (z,{\bf b}) +\bar{a_k}^\dagger
\bar{u_k} ^* (z,{\bf b}) ,
\end{equation}
where the $u_k$'s are normalizable solutions.   Vacuum states can be defined by
\begin{equation} \label{definition of vacua}
a_k|0 \rangle =0 , \ \ \bar{a_k}| \bar{0} \rangle =0 , \ \ \forall k .
\end{equation}
We then have the Bogolubov transformation
\begin{equation} \label{b transformation}
\bar{u} _k = \sum_i \alpha _{ki} u_i + \beta _{ki} u_i ^* , \ \ \bar{a}_k = \sum_i \alpha ^* _{ki} a_i - \beta ^* _{ki} a_i ^\dagger ,
\end{equation}
\begin{equation} \label{b conditions}
\sum_k \alpha_{ik} \alpha_{jk} ^* - \beta_{ik} \beta_{jk} ^* = \delta_{ij} , \ \ \sum_k \alpha_{ik} \beta _{jk} - \beta_{ik} \alpha_{jk} =0.
\end{equation}
When $\beta \neq 0$,  the two vacua are inequivalent and each is an excited state of the other.
For simplicity will assume that the transformation matrices are diagonal, i.e.\ $\alpha_{ij} = \alpha_i \delta_{ij}, \
\ \beta{ij} = \beta_i \delta _{ij}$.  However, the generalization to the non-diagonal case is straightforward, and we will give the necessary
relations for it at the end for completeness.

\subsection{Finding the bulk-boundary propagator with multiple vacua present} \label{bulk-boundary}
A necessary step in the formulation of the AdS/CFT correspondence is the construction of a bulk-boundary propagator $G_{B\partial} ({\bf
b};{\bf b'},z)$ which allows us to write
\begin{equation} \label{bulk boundary gives phi}
\phi(z,{\bf b}) = \int _\partial d{\bf b} \sqrt{-\tilde{g}} G_{B\partial}({\bf b};z,{\bf b'}) \phi_0 ({\bf b})
\end{equation}

Note that in the above equation, $\phi$ is really defined up to a normalizable mode; i.e.,
the bulk-boundary propagator can have arbitrary normalizable modes added on.  Our first step will be to compute a time-ordered Green function for a transition between two, not necessarily equal, vacuum states. Then we will define the bulk-boundary propagator as a particular limit of the bulk-bulk Green function. The time-ordered bulk propagator is a solution to the
equation
\begin{equation} \label{feynman diffeq}
(\Box _x - m^2) G_F(x,x') = -\frac{1}{\sqrt{-g}} \delta^{d+1}(x-x') .
\end{equation}
The boundary conditions on $G_F$ are selected by adding homogeneous solutions of the wave equation to it.

Assume that our
metric has a natural timelike coordinate (not necessarily a Killing vector), with respect to which events can be ordered.  We then have
\begin{eqnarray}\label{feynman two point function definition}
iG_{F\bar{0}{0}} (x,y) &=& \frac{\langle \bar{0} | T(\phi(x) \phi(y) ) | 0 \rangle }{\langle \bar{0} | 0 \rangle } \nonumber \\
&=& \theta(x^0-y^0) \frac{\langle \bar{0} | \phi(x) \phi(y)| 0 \rangle }{\langle \bar{0} | 0 \rangle } +\theta(y^0-x^0) \frac{\langle \bar{0}
| \phi(y) \phi(x)| 0 \rangle }{\langle \bar{0} | 0 \rangle }.
\end{eqnarray}
Note that $\langle \bar{0} | 0 \rangle \neq 1$ in general. Using eqns.\ (\ref{b transformation},\ref{b conditions}), the commutation
relations $[a_i, a^\dagger_j]=\delta_{ij}$, with all others vanishing and the relation \cite{dewitt} $\langle \bar{0}| a_k ^\dagger a_{k'}
^\dagger |0 \rangle=\langle \bar{0}|0 \rangle \beta_{k'} \alpha_k ^{-1} \delta_{kk'}$ we find (for more detail, the reader is directed to
the appendix)
\begin{equation} \label{feynman two point function solution}
iG_{F\bar{0}{0}} (x,y) = \theta(x^0-y^0) \sum_k \bar{u}_k (x) u_k ^* (y) \bigl( \alpha_k ^* - \frac{ |\beta_k| ^2}{\alpha_k} \bigr) +
x\leftrightarrow y.
\end{equation}
In particular
\begin{equation} \label{in in in out bulk}
iG_{F00}= iG_{F\bar{0}{0}} - \sum_k \frac{\beta_k}{\alpha_k} u_k^*(x)u_k^*(y) .
\end{equation}
As expected, the difference between different vacuum choices for $G_F$ is a homogeneous solution.  (We can likewise derive the relation between $G_{F00}$ and $G_{F\bar{0}\bar{0}}$, but we will not do that here.)    This relation is easily generalized to non-diagonal Bogolubov
transformations
\begin{equation}
iG_{F00}= iG_{F\bar{0}{0}} - \sum_{k,k'} \beta_{kk'}\alpha^{-1}_{kk'} u_{k'}^*(x)u_k^*(y) .
\end{equation}
Similar modifications would occur throughout the remainder of this section.
It has been noted by several authors \cite{banks98,vijay99,giddings99} that we can find the bulk-boundary propagator by a simple rescaling
of the time-ordered propagator in the radial direction of AdS.
Following \cite{giddings99} one can easily show in our coordinates that we can write the
bulk-boundary propagator as
\begin{equation}
G_{B\partial \bar{0} 0} ({\bf b}; {\bf b'}, z') = 2 \nu \lim_{z \rightarrow 0} z^{-2h_+} G_{F \bar{0} 0} (x,x') .
\end{equation}

\subsection{Formulation of the correspondence}
With this definition, we can formulate the AdS/CFT correspondence as
\begin{equation} \label{ads cft}
\langle \bar{0} | T \exp i \int_\partial \phi_0 {\cal O} | 0 \rangle_{CFT} = Z_{SUGRA \ \bar{0} 0} (\phi_0) ,
\end{equation}
where the notation on $Z$ implies that we use the $\bar{0} 0$ bulk-boundary propagator for writing $\phi$ in terms of $\phi_0$. We now have a
form of the AdS/CFT correspondence, which takes into account the different boundary conditions at early and late times describing
transitions from various vacua.    The different amplitudes in the CFT are in one-to-one correspondence with the choice of
bulk-boundary propagator in the SUGRA.   In addition, if the vacua are related by a phase (i.e. all the $\beta$s vanish), we have a unique
bulk-boundary propagator and a unique vacuum state, hence the correspondence reduces to the original one. This formulation is particularly
useful if we let $|0 \rangle = |\mathrm{{ in} }\rangle$  and $| \bar{0} \rangle=|\mathrm{{out}}\rangle$ where by $|\rm{in}\rangle$ and $|\rm{out}\rangle$ we mean vacua which are natural for inertial observers
at early at late times.
We can then show that the differences between the vacua are
a result of cosmological particle production.

Let us compute a two point function using the AdS/CFT correspondence to show the effects of having multiple vacua. Consider a scalar field in the
supergravity whose action is given by
\begin{equation} \label{scalar action}
S[\phi] = \frac{1}{2} \int d^d w \, dz \sqrt{-g} (g^{\mu \nu} \partial _\mu \phi \partial _\nu \phi +m^2 \phi^2)  \, ,
\end{equation}
where $z$ is the radial coordinate.
The two point function in the CFT for the operator dual to $\phi$ is
\begin{equation}
\frac{1}{i^2} \frac{\delta ^2 Z_{\bar{0} 0} (\phi_0) } {\delta \phi_0 (x) \delta \phi_0 (y) } |_{\phi_0=0} = \frac{\langle \bar{0} | T({\cal
O}(x) {\cal O} (y) ) | 0 \rangle }{\langle \bar{0} | 0 \rangle } .
\end{equation}
Plugging in for $\phi$ in terms of the bulk to boundary propagator gives
\begin{eqnarray}
\frac{\langle \bar{0} | T({\cal O}(x) {\cal O} (y) ) | 0 \rangle }{\langle \bar{0} | 0 \rangle } = & \ \ \ \ \ \  \ \ \ \ \nonumber \\
&\int
d^d w \, dz  \sqrt{-g} [ g^{\mu \nu} \partial_{\mu} G_{B\partial\bar{0}0} (x;w,z) \partial_\nu
G_{B\partial\bar{0}0} (y;w,z) +m^2 G_{B\partial\bar{0}0} (x;w,z) G_{B\partial\bar{0}0} (y;w,z) ] ,
\end{eqnarray}
Integrating by parts, using the facts that $G_{B \partial}$ satisfies the bulk equations
of motion and that $\lim_{z\rightarrow0} \sqrt{-\tilde{g}} z^{2h_+ - d} G_{B \partial}(x;w,z)=\delta^d (w-x)$ we are left with the boundary
term
\begin{eqnarray}
\frac{\langle \bar{0} | T({\cal O}(x) {\cal O} (y) ) | 0 \rangle }{\langle \bar{0} | 0 \rangle } &=& \lim_{\epsilon\rightarrow0} \int
\epsilon^{1-2h_+} \delta^d(w-x) \partial_z G_{B \partial \bar{0} 0}(y;w,z) |_{z=\epsilon} d^d w \nonumber \\
&=& \lim_{\epsilon\rightarrow0} \epsilon^{1-2h_+} \partial_z G_{B \partial \bar{0} 0}(y;w,z) |_{z=\epsilon} .
\end{eqnarray}
Hence we can conclude that
\begin{eqnarray}
\langle \mathrm{in} | T({\cal O}(x) {\cal O} (y) ) | \mathrm{{ in}} \rangle &=& \frac{\langle \mathrm{{out}} | T({\cal O}(x) {\cal O} (y) ) | \mathrm{{in}} \rangle }{\langle \mathrm{{out}} | \mathrm{in}
\rangle } \nonumber \\
 &+&
2 i \nu \lim_{z\rightarrow 0, \ \epsilon\rightarrow 0} \epsilon^{1-2h_+} \frac{\partial}{\partial z'} \Bigl[
z^{-2h_+} \sum_k \frac{\beta_k}{\alpha_k} u_k ^* (z,b) u_k ^* (z',b') \Bigr]_{z'=\epsilon} ,
\label{master}
\end{eqnarray}
where the $u_k$ are the natural in modes.  Using the scaling as $z \rightarrow 0$ of solutions to the wave equation it is easily shown that
the $z$ dependence drops out, leaving a $k$ dependent multiplicative factor in each term. Thus, this form of the AdS/CFT correspondence is
sensitive to particle creation effects, as we would expect. Similar expressions can be obtained for higher point correlation functions.  If we were computing particle creation directly in the dual, strongly coupled
CFT, we would have to work with a collective field theory for the gauge-invariant composites (see, e.g., \cite{vijay99}). The quantum
numbers of these collective fields would match the quantum numbers of the modes being summed above.   While we have explained the procedure
in asymptotically Poincar\'{e} coordinates, it is simple to generalize to any other asymptotically AdS coordinate system.   In all cases,
the limit will approach the AdS boundary and the the factors of $\epsilon$ and $z$ will be replaced by the appropriate scaling behaviours of
normalizable modes.

In the next two sections we will give two examples of how the above results are used to explore cosmological particle production.

\section{The Schwarzschild-AdS Bubble of Nothing} \label{rossbubble}
It was pointed out in \cite{birm,vijayross} that following \cite{silv02} one can form a ``bubble of nothing'' in AdS by analytically continuing the AdS-Schwarzschild black hole solution. Here we construct the $AdS_4$ analog of the solution found in \cite{birm,vijayross}.   Start with the
solution for the four dimensional AdS-Schwarzschild black hole
\begin{eqnarray}
ds^2 = -f(r)dt^2 + f(r)^{-1} dr^2 + r^2 (d\theta ^2 + \sin^2 \theta d\psi ^2 ) \nonumber \\
f(r) = 1 - \frac{r_0}{r} + \frac{r^2}{l^2} ,
\end{eqnarray}
where $r_0$ is the non-extremality parameter and $l$ sets the length scale.   A double analytic continuation $t\rightarrow i \chi$,
$\theta \rightarrow i \tau + \pi/2$  gives
\begin{equation} \label{schwarz bubble}
ds^2 = f(r) d\chi ^2 + f(r) ^{-1} dr^2 + r^2 ( - d\tau^2 + \cosh^2 \tau d \psi ^2) .
\end{equation}
The space is now cut-off at $r=r_+$ where $r_+$ is the larger solution to
\begin{equation}
r^3 + l^2 r^2 = r_0 .
\end{equation}
To avoid a conical singularity at $r_+$, $\chi$ must have a period of
\begin{equation}
\Delta \chi = \frac{4 \pi l^2 r_+}{3r_+ ^2 + l^2} .
\end{equation}
As in \cite{vijayross}, the boundary metric is de Sitter space times a circle.   Recall that de Sitter space has a family of inequivalent $\alpha$-vacua \cite{dsalpha,bms01}.
  For example, in the global de Sitter coordinates appearing within the final parentheses in (\ref{schwarz bubble}) the in and  out vacua appropriate to early and late times have non-trivial Bogolubov coefficients.   In even dimensional de Sitter spaces
such as the 2d de Sitter factor in (\ref{schwarz bubble}) this leads to cosmological particle production.   It is worth noting that the only
conformally invariant de Sitter vacuum is the Euclidean, or Bunch-Davies vacuum, so this might suggest that this is only valid vacuum state
for the dual CFT \cite{dsalpha}.   However, the presence of the additional circle in (\ref{schwarz bubble}) already provides a scale (see
\cite{vijayross}).  So there is {\it a priori} no reason to discard the other vacua, except in view of various other potential difficulties
with them \cite{dsalpha}.

To investigate particle production and the AdS/CFT correspondence in this
background we must  first solve the wave equation to obtain the mode expansions. In this background the wave equation $\Box \phi - m^2
\phi =0$, becomes
\begin{equation}
-\frac{1}{r^2 \cosh \tau} \partial _\tau (\cosh \tau \partial _\tau \phi) + \frac{1}{r^2} \partial _r ( r^2 f(r) \partial _r \phi) + f(r)
^{-1} \partial_\chi ^2 \phi + \frac{1}{r^2 \cosh ^2 \tau} \partial _\psi ^2 \phi - m^2 \phi = 0 .
\end{equation}
We can separate $\phi$ as $\phi(r,\tau,\chi,\psi) = R(r) T(\tau) e^{i n \psi} e^{i( 2\pi j/\Delta \chi) \chi}$, where $n,j$ are integers. We
define $p = 2 \pi j /\Delta \chi$ and a separation constant $k^2$ to obtain two ordinary differential equations
\begin{eqnarray}
\frac{d}{dr} (r^2 f(r) R'(r)) - (r^2 f(r)^{-1} p^2 + r^2 m^2 -k^2) R(r) = 0 \\
T''(\tau) + \tanh (\tau) T'(\tau) + (k^2 + n^2 /\cosh^2 \tau) T(\tau) = 0 , \label{timeeq}
\end{eqnarray}
where the primes in each equation denote derivatives with respect to the variable of each function.  Note that $k^2$ shows up in the
$T$ equation exactly as a mass would if we were dealing with only the theory on the boundary. The radial equation cannot be solved in closed
form, however we can obtain the asymptotic behavior as
\begin{equation}
R(r) \rightarrow r^{-2h_{\pm}}
\end{equation}
as $r \rightarrow \infty$, where $2h_\pm = d/2 \pm \nu$, and $\nu = 1/2 \sqrt{9+4l^2 m^2}$. The modes that scale as $2h_+$ are normalizable
while the ones that scale as $2h_-$ are non-normalizable.  The radial and angular parts of these solutions will not contribute to Bogolubov transformations relating different vacua because they will have the same form for any basis of solutions.   Thus we only need to focus on the time dependent equation (\ref{timeeq}).

Recall that since de Sitter space has a one complex parameter family of vacua, we could explore the corresponding one parameter family of inequivalent bases of solutions to (\ref{timeeq}).  However, since our interest is simply to illustrate particle production in an AdS/CFT context we will focus on the in and out vacua appropriate to early and late times.  We can then analyze (\ref{timeeq}) by following \cite{bms01} to  arrive at a solvable equation.  We first rewrite the equation in terms of a new coordinate $\sigma= - e ^{2\tau} $
\begin{equation}
\sigma(1-\sigma) T_n ''  + (1/2 - 3/2 \sigma) T_n ' + \biggl[\frac{k^2}{4}\frac{1-\sigma}{\sigma} - \frac{n^2}{1-\sigma} \biggr] T_n =0 ,
\end{equation}
where primes denote differentiation with respect to $\sigma$ and $\mu= \sqrt{k^2-1/4}$. We make the substitution $T^{\mathrm{in}} _n = \cosh ^n \tau
e^{(l+1/2 - i \mu)\tau} f(\sigma) $, where the in superscript has been added in anticipation of this solution describing an incoming wave at
past infinity. The equation now becomes a hypergeometric equation for $f$,
\begin{equation}
\sigma(1-\sigma) f'' + [c-(1+a+b)\sigma]f' - ab f =0 ,
\end{equation}
where the coefficients are given by $a=n+1/2$, $b= n +1/2 - i \mu$, and $c=1- i \mu$. We restrict to the case where $\mu$ is real and
positive. The full incoming solution is then given by
\begin{equation} \label{bubble in solution}
T_n ^{\mathrm{in}} =  \cosh ^n \tau e^{(n+1/2 - i \mu) \tau} F(n+1/2, n+1/2-i \mu; 1-i\mu ; - e^{2 \tau} ) .
\end{equation}
where $F$ is a hypergeometric function and we have chosen not to normalize since it is not important for our purposes. We note that (\ref{s
t equation}) is invariant under time reversal, and the equation is real. Therefore, we have another set of linearly independent solutions
given by $T _n ^{\mathrm{{out}}} (\tau) = T_n ^{in*} (-\tau)$, or
\begin{equation} \label{bubble out solution}
T_n ^{\mathrm{{out}}} =  \cosh ^n \tau e^{-(n+1/2 + i \mu) \tau} F(n+1/2, n+1/2+i \mu; 1+i\mu ; - e^{-2 \tau} ) .
\end{equation}
The asymptotics of these two solutions are given by
\begin{eqnarray} \label{asymptotic bubble}
T^{\mathrm{in}} _n ~\stackrel{ {\tau \rightarrow -\infty}}{\longrightarrow} ~  e^{(1/2 - i \mu) \tau} , \nonumber \\
T^{\mathrm{{out}}}_n ~ \stackrel{\tau \rightarrow \infty}{\longrightarrow} ~ e^{-(1/2 + i \mu) \tau} .
\end{eqnarray}
Thus, our solutions are good candidates for defining in and out vacua respectively.

The radial and angular parts of the wavefunction will clearly not play a role in the Bogolubov coefficients between the in and out states. Using the general relation between hypergeometric functions \cite{as}
\begin{eqnarray}
F(a,b;c;z) = \frac{ \Gamma(c) \Gamma(b-a)}{ \Gamma(c-a) \Gamma(b)} (-z)^{-a} F(a,a+1-c;a+1-b;1/z) + \nonumber \\
\frac{\Gamma(c) \Gamma(a-b)} {\Gamma(a) \Gamma(c-b)} (-z)^{-b} F(b,b+1-c;b+1-a;1/z) .
\end{eqnarray}
we may compute the relation between $T^{\mathrm{in}} _n$ and $T^{\mathrm{{out}}} _n$ and find
\begin{equation}
T^{\mathrm{in}} _n = \alpha _n T ^{\mathrm{{out}}} _n + \beta_n T^{\mathrm{{out}}*} _n\,
\end{equation}
with
\begin{equation}
\alpha_n= \frac{\Gamma(1-i\mu)\Gamma(-i\mu)}{\Gamma(1/2-n-i\mu)\Gamma(1/2+n-i\mu)} , \ \ \ \beta_n=-i \frac{(-1)^n}{\sinh(\pi \mu)} .
\end{equation}
The number of particles in the in vacuum for the $n$-th mode is given by
\begin{equation}
\langle \rm{in}| N_n | \mathrm{\rm{in}} \rangle = |\beta_n|^2 = \frac{1}{\sinh ^2 (\pi \mu) } .
\end{equation}
Because this is independent of $n$ and of $j$ (the other angular momentum) there will be an infinite amount of particle production at high momenta.  So strictly speaking we should be cutting off our in and out vacua suitably at high momenta, but we will not delve into this here.

How is this particle production by the cosmological expansion of the spacetime reflected in the dual CFT?
From (\ref{master}), we can see that the  $\langle \mathrm{in}| \cdots |\mathrm{in} \rangle $ correlation function differs from the $\langle \mathrm{out} | \cdots |\mathrm{in} \rangle $ by a particle creation term.  For the bubble of nothing spacetime that we are considering in this section, this term is given by the expression
\begin{equation}
\sum_{n,j}
2 i \nu \Bigg\{ \lim_{r \rightarrow \infty,  L \rightarrow \infty} L^{-1+2h_+} r^{2h_{+}} R(r) \frac{\partial R(r')}{\partial r'} \Bigg\}
\frac{\beta_n}{\alpha_n} e^{-n(\psi + \psi')}e^{-i\Bigl({\frac{2\pi j} {\Delta \chi}}(\chi+\chi')\Bigr)}T^*_n(\tau)T^*_n(\tau')
\end{equation}
The $r$ dependence in the above expression cancels out from the term in the braces which just yields $2h_+$. Also, the sum over $j$  gives a term proportional to $\delta(\chi+\chi')$. The summation over $n$ is non-trivial to carry out explicitly and we will not do so here.

\section{The bubbling boundary}
\label{Schwars}

As another example we turn to the spacetime formed by replacing the flat part of the metric of $AdS_5$ in Poincare
coordinates with the double analytically continued Schwarzschild bubble of nothing found in \cite{silv02}. The metric is
\begin{equation}
\label{s bubble metric} ds^2=\frac{d \rho^2}{\rho^2} + \rho^2 \biggl( -r^2 d \tau^2 + \frac{r^2dr^2}{r^2-2Mr} + \frac{r^2-2Mr}{r^2} {d
\chi^2} + r^2 \cosh ^2 \tau d \theta^2 \biggr) .
\end{equation}
Note in this section we have changed notation slightly, $\rho$ ($=1/z$) now represents the radial variable in the bulk, and $r$ the radial
variable on the boundary. The radial variable on the boundary, $r$, is restricted to $r\geq 2M$. In order for the spacetime to be regular at
$r=2M$ we require $\chi$ to be periodic with period $8 \pi M$, as in the standard Euclidean black hold background, while $\theta$ has the
normal $2 \pi$ periodicity. This spacetime has a mild curvature singularity at the horizon, due to the square of the Riemann tensor
diverging there. This singularity is similar to the one found in the AdS-black string solution \cite{chamblin99}. It has recently been
debated whether an instability in the black string case leads to a resolution of the singularity by the black string pinching off to form a
cigar or pancake like shape \cite{gregory00,gibbons02}. In our case, the possible instability will be different, as our time and angular
coordinates are different, hence the possible mode will {\it not} simply be an analytic continuation of the black string mode. In addition,
because the space actually ends at the bubble, proper boundary conditions must be imposed which will be different from the black string
case. The stability and endpoint of this background is an interesting problem in its own right. Here, we merely point it out since our main
interest is to use it as a simple example.

The scalar wave equation separates into a radial and a boundary part, and we begin by examining the latter, which is $\tilde{\Box} \psi (\tilde{x}) - k^2 \psi
(\tilde{x}) =0$, where tildes refer to the boundary metric only.  (The bulk radial equation will not play a role in the Bogolubov transformations.)    In detail, this is
\begin{equation} \label{boundary s bubble wave equation}
- \frac{1}{r^2 \cosh \tau} \partial _\tau (\cosh \tau \partial _\tau \psi) + \frac{1}{r^2} \partial _r \biggl( r^2 \biggl( 1 - \frac{2M}{r}
\biggr) \partial _r \psi \biggr) + \biggl (1 - \frac{2M}{r} \biggr) ^{-1} \partial ^2 _\chi \psi + \frac{1}{r^2 \cosh ^2 \tau} \partial
_\theta ^2 \psi - k^2 \psi = 0.
\end{equation}
Since $\partial _\chi$ and $\partial _\theta$ are Killing vectors, we choose as our trial ansatz
\begin{equation} \label{s ansatz}
\psi (\tau, r , \chi, \theta) = T(\tau) R(r) \cos(n\chi / 4M) \cos(l \theta) ,
\end{equation}
where the arguments in the $\chi$ and $\theta$ parts are determined by their respective periodicity requirements, with $n$ and $l$ taking
integer values. Note also that we have chosen a basis of functions in these variables that are real (we could just as easily put in sine
functions, or a linear combination of cosines and sines, our conclusions will be the same). We are thus reduced to two separate ordinary
differential equations
\begin{equation} \label{s t equation}
\frac{1}{ \cosh \tau} \frac{d}{d \tau} \biggl( \cosh \tau \frac{dT}{d \tau} \biggr) +\biggl(p^2 + \frac{l^2}{\cosh ^2 \tau} \biggr)T =0 ,
\end{equation}
\begin{equation} \label{s r equation}
\frac{d}{dr} \biggl( r^2 \biggl( 1- \frac{2M}{r} \biggr) \frac{dR}{dr} \biggr) + \biggl(- r^2 \biggl[k^2 + \biggl(1-\frac{2M}{r}\biggr)^{-1}
\frac{n^2}{16 M^2}\biggr] +p^2 \biggr) R =0 ,
\end{equation}
where $-p^2$ is a constant resulting from the separation of variables. We first focus on the radial equation. This equation is difficult to
solve analytically, and is almost identical to the normal radial Schwarzschild equation (for a discussion of these see
\cite{page76,jensen86,bd} and references therein). We only need that the decaying asymptotic for large $r$ ($r\gg 2M$) is given by
\begin{eqnarray} \label{s r asymptotic}
R \rightarrow  \frac{K_\nu (\sqrt{k^2+n^2/16M^2} r )}{\sqrt{r}} , \ \ \ \ n, k \neq 0 \\
R \rightarrow r ^{- 1/2 \pm  \nu} , \ \ \ \ \ \ n,k =0 .
\end{eqnarray}
where $\nu = \sqrt{1/4 - p^2}$.   Fortunately, a detailed form of the radial solution is not necessary for computing the Bogolubov
transformations between in and out vacua.

The time dependent equation (\ref{s t equation}) is identical to (\ref{timeeq}) in Sec.~\ref{rossbubble}.   Therefore, solving it in the same way we find
\begin{eqnarray}
T_l ^{\mathrm{in}} &=&  \cosh ^l \tau e^{(l+1/2 - i \mu) \tau} F(l+1/2, l+1/2-i \mu; 1-i\mu ; - e^{2 \tau} ) ,
\label{s in solution} \\
T_l ^{\mathrm{{out}}} &=&  \cosh ^l \tau e^{-(l+1/2 + i \mu) \tau} F(l+1/2, l+1/2+i \mu; 1+i\mu ; - e^{-2 \tau} ) \, ,
\label{s out solution}
\end{eqnarray}
where $\mu=\sqrt{p^2-1/4}$. As before the asymptotics of these solutions make them good candidates for defining in and out vacua
respectively. Then the Bogolubov transformation between the in and out vacua is $T^{\mathrm{in}} _l   = \alpha_l T^{\mathrm{{out}}} _l +
\beta_l T^{\mathrm{{out}}*} _l$ where the coefficients are given by
\begin{equation}
\alpha_l= \frac{\Gamma(1-i\mu)\Gamma(-i\mu)}{\Gamma(1/2-l-i\mu)\Gamma(1/2+l-i\mu)} , \ \ \ \beta_l=-i \frac{(-1)^l}{\sinh(\pi \mu)} .
\end{equation}
So the number of out particles in the in vacuum for the $pln$-th mode is given by
\begin{equation}
\langle \rm{in}| N_{pln} | \mathrm{\rm{in}} \rangle = |\beta_l|^2 = \frac{1}{\sinh ^2 (\pi \mu) } .
\end{equation}
Because this is independent of the momentum quantum numbers there is an infinite amount of particle production.  (Note that this does not agree with the geometric optics approximation in \cite{silv02}.)

Using (\ref{master}) as in the previous section translates this bulk particle production due to cosmological expansion of the solution into
particle production in the dual field theory.  Notice that $k$, which is the radial momentum eigenvalue in the AdS bulk acts as a ``mass''
as far as the boundary coordinates are concerned.  Thus, the sum over AdS radial momenta that appears in the expression for the total number
of particles produced will be related in the CFT to a sum over different mass shells that will appear in the spectral decomposition of a
large $N$ collective field for the dual operator.  (See \cite{vijay99} for related  comments.)

\section{Conclusions and Future Directions} \label{conclusions}
We have shown how to formulate the AdS/CFT correspondence when time dependence leads to multiple vacua and particle production. Using this
formulation we showed how two point correlators in the CFT can be computed in various vacua from the bulk point of view. We have also found a
simple relation between particle production in the bulk and the boundary and illustrated these facts in examples.  It is easy to extend our comments to higher point correlators.  In general, the need to avoid poles in various ways will provide a role for real time thermal field theory.   It would be of interest to apply our analysis in settings that offer insights into the behavior of strongly coupled theories in time dependent backgrounds.

\appendix
\section{Green's function computation}
The computation of the propagator can be reduced to computing $\langle \bar{0} | \phi(x) \phi (x') | 0 \rangle$. Using the definitions of
the vacua and the mode expansions we have
\begin{eqnarray}
\langle \bar{0} | \phi(x) \phi (x') | 0 \rangle= \sum_{k,k'} \langle \bar{0} | (\bar{a}_k \bar{u}_k(x)+ \rm{h.c.})(a_{k'} u_{k'}(x)
+\rm{h.c.})|0\rangle
\nonumber \\
=\sum_{k,k'} \langle \bar{0} | \bar{u}_k(x) u_{k'}^*(x') \bar{a}_k a_{k'} ^\dagger|0\rangle .
\end{eqnarray}
Using the Bogolubov transformations (\ref{b transformation}) we get
\begin{eqnarray}
\langle \bar{0} | \phi(x) \phi (x') | 0 \rangle=\sum_{k,k'} \bar{u}_k(x) u_{k'}^* (x') \langle\bar{0} | (\alpha_k ^*a_k -\beta_k ^* a_k
^\dagger) a_{k'} ^\dagger|0\rangle ,
\end{eqnarray}
but we can use the commutation relations, $[a_i,a_j^\dagger]=\delta_{ij}$ to compute the first term, and the relation (see \cite{dewitt} for
the derivation) $\langle \bar{0}| a_k ^\dagger a_{k'} ^\dagger |0 \rangle=\langle \bar{0}|0 \rangle \beta_{k'} \alpha_k ^{-1} \delta_{kk'} $
for the second term, which gives
\begin{eqnarray}
\langle \bar{0}| a_k ^\dagger a_{k'} ^\dagger |0 \rangle= \langle \bar{0}|0\rangle \sum_k \bar{u}_k(x) u_k^* (x') \Bigl( \alpha_k^* -
\frac{|\beta_k|^2}{\alpha_k} \Bigr) .
\end{eqnarray}
Plugging this relation into the definition of the time-ordered propagator gives (\ref{feynman two point function solution}).

\begin{acknowledgments}
We thank Jan de Boer, Esko Keski-Vakkuri, Djordje Minic and Simon Ross for useful comments.
Work on this project was supported by the DOE under cooperative research agreement DE-FG02-95ER40893, and by an NSF Focused Research Grant DMS0139799 for ``The Geometry of
Superstrings".
\end{acknowledgments}

\end{document}